# Consideration of prior information in the inference for the upper bound earthquake magnitude

Mathias Raschke, Freelancer, Stolze-Schrey-Str. 1, 65195 Wiesbaden, Germany, E-Mail: mathiasraschke@t-online.de

Abstract: The upper bound earthquake magnitude (maximum possible magnitude) of a truncated Gutenberg-Richter relation is the right truncation point (right end-point) of a truncated exponential distribution and is important in the probabilistic seismic hazard analysis. It is frequently estimated by the Bayesian inference. This is a non-regular case and suffers some shortcomings in contrast to the Bayesian inference for the regular case of likelihood function. Here previous non-Bayesian inference methods are outlined and discussed as alternatives including, the formulation of the corresponding confidence distributions (confidence or credible interval). Furthermore, the consideration of prior information is extended to non-Bayesian estimation methods. In addition, two new estimation approaches with prior information are developed. The performances of previous and new estimation methods and corresponding confidence distributions were studied by using Monte Carlo simulations. The mean squared error and bias were used as the main performance measures for the point estimations.
In summary, previous, and new alternatives overcome the fundamental weakness of the Bayesian inference for the upper bound magnitude. These alternatives are not perfect but extend the opportunities for the estimation of the upper bound magnitude considerably, especially the consideration of prior information. A prior distribution reduces the global mean squared error but leads to a local bias – small upper bound magnitudes are overestimated, and high upper bounds are underestimated.

## 1 Introduction

The maximum possible earthquake magnitude $m_{max}$ is the upper bound of the well-known Gutenberg-Richter relation, which forms a truncated exponential distribution with a defined left threshold magnitude. Page (1968) and Cornell and Vanmarke (1969) suggested the right truncation, which is realized in most subsequent distribution models for magnitudes (e.g. Cosentino et al. 1977, Utsu 1999, Wössner et al. 2015). The probability density function (PDF) is the first derivative of the cumulative distribution function (CDF). The CDF of the truncated exponential distribution for magnitudes (also called doubly truncated exponential distribution) is:

$$F(x) = \frac{1-\exp(-\beta(x-m_{min}))}{1-\exp(-\beta(m_{max}-m_{min}))}, \beta > 0, m_{min} \leq x \leq m_{max}. \tag{1}$$

Correspondingly, the statistical estimation of the upper bound magnitude $m_{max}$ was the topic of various papers. These inference methods were applied in numerous models. The approach by Cornell (1994) is very popular and considers the Bayesian method with actual prior information for the estimation of $m_{max}$. Some weaknesses of this approach were discussed already (Holschneider et al. 2011). Kijko (2012) pointed out that this method still includes a bias and it is the non-regular case of the maximum likelihood function (cf. Smith 1985). However, the Bayesian inference was recently used for the parametrization of a number of important probabilistic seismic hazard analyses (PSHA), such as by Wössner et al. (2015), EPRI (2015) and Grünthal (2018). This explains the motivation for the current paper, where we research, discuss, and extend the Bayesian approach as well as validate the corresponding performance. The main issue is the point estimation of $m_{max}$ including the corresponding confidence/credible intervals. Previous contributions and researches are also discussed in different parts of the paper. In the following section 2, the classical Bayesian estimation is explained and the differences between the regular case and the estimation of $m_{max}$ is presented in detail; multiple weaknesses and issues are explained. In section 3, important non-Bayesian alternative inference methods are outlined and discussed. And the consideration of prior information is extended to non-Bayesian estimation methods. In addition, a new inference approach is formulated. A concept for the performance evaluation is formulated and applied in section 4. Finally, the results are discussed and concluded.

A number of aspects of the inference for the upper bound magnitude $m_{max}$ are researched and discussed here, but not all topics can be considered. Uncertainties of the magnitude measurement lead to a biased estimation of the Gutenberg-Richter relation (Tinti and Mulargia 1985; Rhoades and Dowrick 2000), which means that the measurements of the single events are biased, including the maximum observed magnitude. This issue will not be considered in the current study.

Campbell (1982), Holschneider et al. (2011) and Zöller and Holschneider (2016) focused on the distribution of the maximum observed magnitude of a period. They also applied Bayesian inference and calculated the predictive distribution (cf. Beirlant et al. 2004) even though they used a different term. Holschneider et al. (2011) called it confidence interval. Usually, and in this paper too, the confidence interval refers to an estimated parameter and not to a random variable. Campbell (1982), called the predictive distribution as Bayesian distribution. It is highlighted here that the predictive distribution is not part of the current research and includes serious shortcomings according to Raschke (2017b).

Furthermore, the author draws attention to the fact that only independent and identically distributed random variables, here the magnitudes, are considered in the following study.

## 2 Classical Bayesian inference for the upper bound magnitude and corresponding weaknesses

### 2.1 The Approach

The classical Bayesian estimation method is conventionally applied in seismology (c.f. Johnston et al. 1994, Grünthal 2004). Therein, prior information is used, presented by a prior distribution $\pi(\theta)$ of parameter θ. Furthermore, the likelihood function

$$f(\mathbf{x};\theta) = \prod_i^n f(x_i;\theta) \qquad (2)$$

is applied for the sample with $n$ observations $x_i$. The posterior distribution is the normalized product of the prior distribution and likelihood function (Lindsey 1996, DasGupta 2008)

$$\pi(\theta|\mathbf{x}) = \frac{\pi(\theta)f(\mathbf{x};\theta)}{\int_{\theta^*} \pi(\theta^*)f(\mathbf{x};\theta^*))d\theta^*}. \qquad (3)$$

The point estimation is usually the expectation of the posterior distribution

$$\hat{\theta} = E[\theta|\mathbf{x}] = \int_\theta \theta \pi(\theta|\mathbf{x})d\theta. \qquad (4)$$

The median of the posterior is another point estimator. The credible interval (credible region) is given by the quantiles of the posterior distribution. The credible interval has the same meaning as the confidence interval (confidence region) of the non-Bayesian inference. Therefore, only the term confidence interval is used and the corresponding distribution (the posterior in case of Bayesian inference) is called confidence distribution here with CDF $\hat{F}_{con}$. Additionally, the term 'Bayesian' is only used for inference methods with prior information about the estimated parameter and the maximum likelihood function.

There exists a special situation in case of the upper bound magnitude of the truncated exponential distribution. The likelihood function for parameter $m_{max}$ can be reduced to

$$f(m_{max}) = \begin{matrix} 0 \; if \; x < m_{max}^{obs} \\ \alpha(1 - \exp(-\beta(m_{max} - m_{min})))^{-n} \end{matrix} \qquad (5)$$

and is determined by the sample maximum, being the maximum observed magnitude $m_{max}^{obs}$ and the sample size $n$ being the number of observed magnitude higher than the threshold $m_{min}$ (cf. Cornell 1994, Grünthal et al. 2004). Equation (5) has the global maximum at $m_{max}^{obs}$ and converges to a constant $\alpha$ with increasing $m_{max}$. Normalized examples ($\alpha = 1$) are shown in Figure 1. The variants demonstrate that the influence of the observation threshold $m_{min}$ is negligible since the sample size $n$ decreases by increasing $m_{min}$.

The application of the Bayesian methods to the upper bound magnitude is not a regular case (Smith 1985). The integration of the likelihood function over the entire parameter scale would result in a finite number in the regular case; it should be a measurable function for the classical Bayesian inference (Smith 1998, van der Vaart 1998, p. 140). However, the integration of (2) has an infinite result in case of the truncated exponential distribution. This issue was already stated by Holschneider et al. (2001). The specific behaviour of the estimator is explained in the following section.

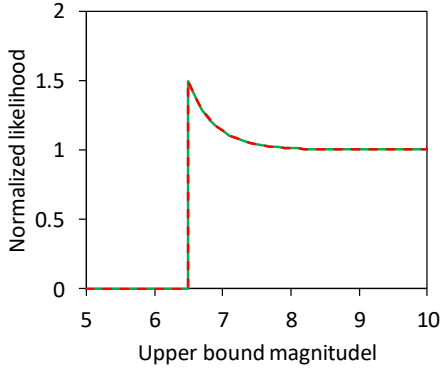

*Figure 1: Examples of likelihood functions according to equation (5) with $\beta = ln(10)$ and $m_{max}^{obs} = 6.5$ for a sample size $n = 4000$ and $m_{min} = 2.5$ (green solid line) and size $n = 400$ and $m_{min} = 3.5$ (red broken line).*

## 2.2 The unusual dependents of Bayesian point estimation for the upper bound magnitude on the information content

The result of the classical Bayesian inference depends on the standard deviation of the prior distribution. It contains a maximum of information if its standard deviation is 0 and loses all information if its standard deviation converges to infinity. The Bayesian point estimation (4) equals the expectation of the prior distribution in the case of maximum information. It also equals the ML point estimation in the case of no information due to an infinite standard deviation of the prior. The estimation of the expectation of a normal distributed random variable $X$ with a PDF

$$f(x) = \frac{1}{\sigma\sqrt{\pi}} e^{-\frac{1}{2}\left(\frac{x-\mu}{2}\right)^2} \tag{6}$$

illustrates this behaviour. The known variance is $V[X] = \sigma^2 = 1$. The sample size is $n = 10$ and sample mean is $\bar{x} = 1.5$; the prior information is also normal distributed with expectation $E[X] = \mu = 0.25$. The dependence of the Bayesian point estimation on the standard deviation of the prior distribution is shown in Figure 2a. The ML point estimation is equal to the sample mean $\bar{x} = 1.5$.

The Bayesian point estimation for the upper bound magnitude does not have such a reasonable behaviour. Although the Bayesian point estimation equals the expectation of the prior in case of a standard deviation = 0 of the prior distribution, it converges to infinity with increasing standard deviation and not to the ML point estimation. The latter is $m_{max}^{obs}$. An example is shown in Figure 2b. The sample size is $n = 4000$ for $m_{min} = 2.5$, $m_{max}^{obs} = 5.5$ and $b = 1$. The normal prior distribution has an expectation $\mu = 6.5$.

The limit behaviour of the Bayesian estimation for the upper bound magnitude is non-acceptable from a statistical point of view. Moreover, the relation has a global minimum in contrast to the graph of the estimations for the normal distribution in Figure 2a. Further issues of the Bayesian inference are explained in the flowing two sections.

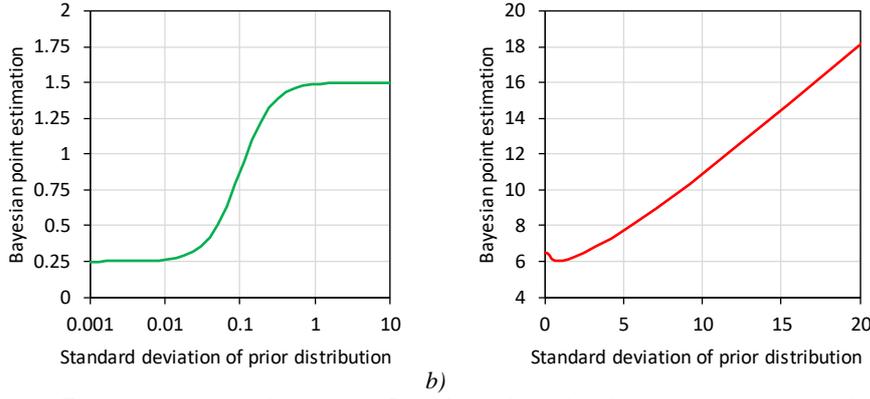

*a)* *b)*

*Figure 2: Examples of the relation between Bayesian point estimation and the standard deviation of the normal prior distribution: a) for the expectation of a normal distribution, b) for an upper bound magnitude.*

## 2.3 The influence of the type of prior distribution

A further issue of the classical Bayesian inference for upper bound magnitude is the influence of the distribution assumption for the prior information. Coppersmith (1994) suggested a normal distribution as prior without any validation or test of the assumption. It can be easily shown by an example, that the influence of the distribution assumption on the estimation result is not negligible. Therein, the entire sample size is $n = 4000$ for $m_{min} = 2.5$ and $b = 1$. The prior distribution has an expectation of 6.5, the standard deviation is 0.75. Three types of prior distributions are considered according to Figure 3b, a normal distribution, a log-normal distribution, and a symmetric beta distribution (Johnson 1994 and 1995; formulations are presented in the appendix). The beta distribution differs less from the normal distribution than the log-normal distribution does. The corresponding graphs for the relations between $m_{max}^{obs}$ and $\hat{m}_{max}$ are presented in Figure 3b. Differences of 0.2 are not rare and not the maximum value.

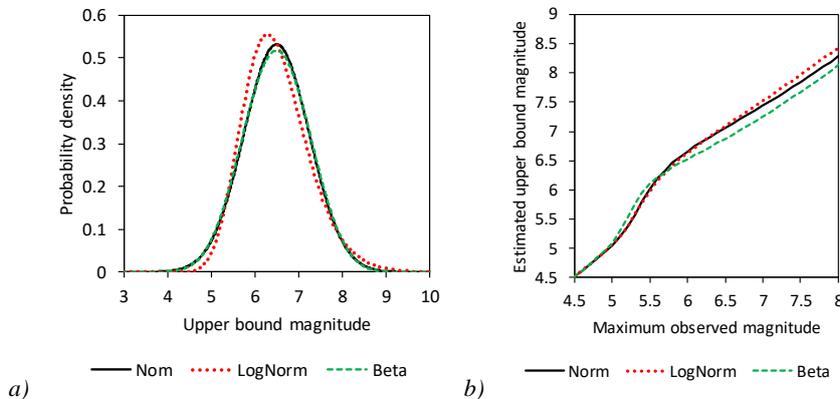

*a)* *b)*

*Figure 3: Influence of the type of prior distribution: a) PDFs of considered prior distributions with the same expectation and standard deviation, b) relation between maximum observed magnitude and the Bayesian point estimation.*

## 2.4 Propagation of uncertainties in the prior information

The general prior information for the estimation of an upper bound magnitude are prior estimations, which implies uncertainties and errors (Tinti and Mulargia 1985; Rhoades and Dowrick 2000).

Here, the consequences are demonstrated for an example. The sample size is $n = 2000$, $m_{min} = 2.5$ and $\beta = \ln(10)$. The correct normal prior distribution has an expectation of 6.5 and the standard deviation of 0.75. The assumed errors are 0.25, one for the expectation and one for the standard deviation of the prior. The resulting graphs for the relation between maximum observed magnitude and the Bayesian point estimation are presented in Figure 4. The influence of the error of the false expectation of the prior is higher than of the influence of the error of the standard deviation of the prior in the range of small and medium maximum observed magnitudes. The maximum of the differences between the correct point estimation and the estimation with a bias is above 0.2.

These errors of prior information can also be interpreted as biases since the errors of a prior distribution are static and influence every estimation based on this prior distribution.

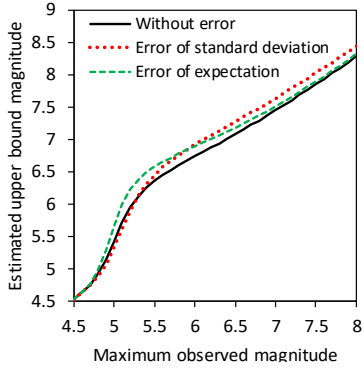

*Figure 4: Influence of errors in the prior information.*

# 3 Alternative inference methods for the upper bound magnitude

## 3.1 Non-Bayesian inference methods

There are different non-Bayesian estimation methods for the upper bound magnitude. Some important methods are outlined here. The simplest estimator is the ML estimation (Robson and Whitlock 1964)

$$\hat{m}_{max} = m_{max}^{obs}. \tag{7}$$

This point estimation is biased. Many estimators correct the bias with

$$\hat{m}_{max} = m_{max}^{obs} + \hat{\Delta} \tag{8}$$

The simplest correction method is to add the difference between the largest and the second largest observation with

$$\hat{\Delta} = m_{max}^{obs} - m_{2nd\ max}^{obs}. \tag{9}$$

Further estimators for $\hat{\Delta}$ are published by Beg (1982) and Hannon and Dahiya (1999). One well-known estimation method is based on the formulation by Kijko and Graham (1998) and Kijko (2004)

$$\hat{m}_{max} = m_{max}^{obs} + \int_{m_{min}}^{\hat{m}_{max}} [F(m)]^n dm. \tag{10}$$

where $F$ is the CDF of the magnitude, i.e. the truncated exponential distribution, and $n$ is the sample size. The theoretical background of the method was already provided by Cooke (1979) and Kijko (1983). According to Kijko (2004), the equation is solved for $\hat{m}_{max}$ iteratively with the start value $\hat{m}_{max} = m_{max}^{obs}$. One major issue is that one solution for equation (10) results in the untruncated case with

$$\hat{m}_{max} = \infty. \tag{11}$$

The main questions are: Is there a further, finite solution for equation (10), and how can this be proven? A brief numerical research suggests that no finite solution for equation (10) exists in any case (Figure 5). The sample size should not influence this result since the sample size is also determined by the observation threshold $m_{min}$ (cf. Figure 1). The issue can be omitted if equation (10) is applied without iteration; integration limit $\hat{m}_{max}$ could be replaced by $m_{max}^{obs}$. The current concern points in the same direction as the comments by Zöller (2017). The author also assumes that a similar problem exists with the estimator of Hannon and Dahiya (1999) and would also apply to the following method if this was also computed iteratively.

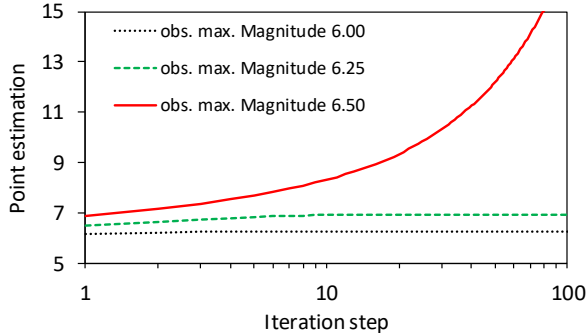

*Figure 5: Iterative solutions for (8) for a truncated exponential distribution with $m_{min} = 5.5$ and $\beta = \ln(10)$ and sample size $n = 5$.*

This estimator is also popular in the seismological community. It is the bias free method, which considers the sample size $n$ and the PDF $f$ of the untruncated magnitude distribution with

$$\hat{\Delta} = \frac{1}{n f(m_{max}^{obs})}. \tag{12}$$

The corresponding standard error is also $\hat{\Delta}$. Pisarenko (1991) has introduced this method in seismology, which was published at first by Tate (1959).

In the estimation for $\hat{m}_{max}$ in equation (8), some elements of the concrete estimators are replaced, such as the sample size by the product of observation period and the pre-estimated seismicity rate. The influence of uncertainties of pre-estimated parameters has been considered by Pisarenko et al. (1996), Kijko and Graham (1998) and Kijko (2004) in special variants for their estimation methods. They used the term Bayesian for this modification. However, these are not classical Bayesian inference methods since no prior information on $m_{max}$ is processed. If the author understands their Bayesian variants correctly, then they used the concept of predicting distribution (Raschke 2017b, Raschke 2018). This is the mixture of distributions (cf. Das Gupta 2008, chapter 22).

The influence of errors of pre-estimated parameters such as the $\beta$-value can be quantified by the delta method, also called Taylor approximation, for error propagation (Raschke 2014a, 2017a).

The classical methods of extreme value statistics for a finite right end-point estimation have also been applied for the maximum earthquake magnitude by Pisarenko et al. (2008, 2014). These methods do not perform well, as explained by Raschke (2015). The convergence of the tail of a truncated exponential distribution to a Generalized Pareto distribution of the Weibull domain is too slow. The underlying models assumed in extreme value methods are more general than the specific truncated exponential model and hence it is quite natural that the properties of those estimators are less good than the methods which are based on the truncated exponential model.

The limitation of classical extreme value statistics in cases of truncated distribution tails was overcome by Beirlant et al. (2016, 2017). However, the performance of their method is not very high. Previous methods have a lower mean squared error (MSE) according to Beirlant et al. (2018). The latter publication also offers a good overview about different estimation methods. In addition, Vermeulen and Kijko (2017) established that the estimation according to equations (10) and (12) is asymptotical equivalent. This means that the difference of the point estimations converges to 0 with increasing sample size and fixed threshold $m_{min}$. Once again, this method is more universal than the special estimation methods for the truncated exponential distribution.

Another issue in the estimation of the upper bound magnitude is the formulation of the confidence intervals. This is not formulated for equation (12) and other estimators according to equation (8). However, the estimation errors of $\hat{\Delta}$ can be quantified for equation (12) and are equal to $\hat{\Delta}$. This matches the characteristics of an exponential distribution, where its standard deviation equals its expectation. It also matches with the fact that the difference of $m_{max} - m_{max}^{obs}$ is exponentially distributed (Hannon Dahiya 1999). Therefore, the confidence interval might be reasonably approximated by an exponential distribution.

Pisarenko (1991) has also formulated a confidence distribution $\hat{F}_{con}$ with the truncated distribution $F$ of the magnitudes (here the truncated exponential distribution)

$$\hat{F}_{con}(x) = 1 - \left[F(m_{max}^{obs}; m_{max} = x)\right]^n. \tag{13}$$

The original argument of $F$ is $m_{max}^{obs}$ where $m_{max}$ is the truncation parameter. In $\hat{F}_{con}$, $m_{max}$ is the argument and $m_{max}^{obs}$ is a parameter. Equation (13) does not provide a simple point estimation since the distribution has probability mass for $m_{max} = \infty$. This was already noted by Kijko (2004). Raschke (2012) overcame this issue by applying a normalization. He formulated with the CDF $F^*$ (here the untruncated exponential distribution):

$$\hat{F}_{con}(x) = \begin{cases} 0 & \text{if } x \leq m_{max}^{obs} \\ \frac{1 - [F^*(m_{max}^{obs})/F^*(x)]^n}{1 - [F^*(m_{max}^{obs})]^n} & \text{if } x > m_{max}^{obs} \end{cases}. \tag{14}$$

The new confidence distribution for $m_{max}$ has an expectation, which is the point estimation and is computed numerically. The method is called constructed estimator. The normalization is basically the same as the consideration of a flat, information free prior distribution in the Bayesian inference. The poster contribution of Kijko and Smit (2014) went even further. They suggested combining an informative Bayesian prior distribution with the PDF of a confidence distribution. Alas, they did not provide a mathematical validation of their concept. Here, the approach of Raschke (2012) and the idea of Kijko and Smit (2014) is extended and a mathematical justification is presented in the following section.

## 3.2 Alternative consideration of prior information in non-Bayesian inference

The inference for a parameter $\theta$ according to the classical Bayesian procedure was already explained in section 2.1 with equations (2-4). The Bayesian inference can be also applied without actual prior information. A flat prior distribution does not contain information. This is a uniform distribution with homogenous probability density $1/(2a)$ for the interval $[-a, a]$ and 0 probability density outside the interval. The Bayesian formulation is then

$$\pi(\theta|\mathbf{x}) = \begin{matrix} \frac{\frac{1}{2a}f(\mathbf{x};\theta)}{\int_{-a}^{a} \frac{1}{2a}f(\mathbf{x};\theta^*))d\theta^*} \text{ if } |\theta| \leq a \\ 0 \text{ if } |\theta| > a \end{matrix}. \qquad (15)$$

what can be simplified to

$$\pi(\theta|\mathbf{x}) = \begin{matrix} \frac{f(\mathbf{x};\theta)}{\int_{-a}^{a} f(\mathbf{x};\theta^*))d\theta^*} \text{ if } |\theta| \leq a \\ 0 \text{ if } |\theta| > a \end{matrix}. \qquad (16)$$

Now, the bound $a$ converges to infinity and the Bayesian estimation becomes

$$\pi(\theta|\mathbf{x}) = \frac{f(\mathbf{x};\theta)}{\int_{-\infty}^{\infty} f(\mathbf{x};\theta^*))d\theta^*}. \qquad (17)$$

It is highlighted again, that (17) cannot be applied to the non-regular case. The expectation of a normal distribution was already used as an example for the Bayesian inference in section 2.2. It is also applied here for the Bayesian estimation of the expectation of a normal distribution with known variance and flat prior distribution. The actual distribution has expectation $\mu = 3$ and variance $\sigma^2 = 1$ and the sample size is $n = 10$. Details are presented in Figure 6a. The posterior confidence distribution of the Bayesian inference is shown in Figure 6b. It equals the non-Bayesian confidence distribution with the sample mean as expectation and variance $\sigma_{con}^2 = \sigma^2/n$.

There is a simple extension of the Bayesian inference – the Bayesian updating (Jaffray 1993, Zhang and Mahadevan 2000): A 1st Bayesian was computed for sample $\mathbf{x}$. The result is the 1st posterior distribution $\pi(\theta|\mathbf{x})$. Now, there are further observations $\mathbf{x}_*$, which should be considered. This done by an update. Therein, $\pi(\theta|\mathbf{x})$ is the 2nd prior information, and the result is the 2nd posterior distribution $\pi_*(\theta|\mathbf{x}, \mathbf{x}_*)$ with

$$\pi_*(\theta|\mathbf{x}, \mathbf{x}_*) = \frac{\pi(\theta|\mathbf{x})f(\mathbf{x}_*;\theta)}{\int_{\theta^*} \pi(\theta|\mathbf{x})f(\mathbf{x}_*;\theta)d\theta^*}. \qquad (18)$$

Now the case is considered, that there is already a Bayesian posterior confidence distribution $\pi(\theta|\mathbf{x})$, and new information are provided by the 2nd prior distribution $\pi_*(\theta)$. The estimation can be updated similar to (18) and a 2nd posterior confidence distribution $\pi_*(\theta|\mathbf{x})$ is generated with

$$\pi_*(\theta|\mathbf{x}) = \frac{\pi_*(\theta)\pi(\theta|\mathbf{x})}{\int_{\theta^*} \pi_*(\theta^*)\pi(\theta^*|\mathbf{x}))d\theta^*}. \qquad (19)$$

If the 1st posterior $\pi(\theta|\mathbf{x})$ is generated by a flat prior distribution according to equation (15-17), then $\pi(\theta|\mathbf{x})$ can be replaced in (19) by (17) and we get

$$\pi_*(\theta|\mathbf{x}) = \frac{\pi_*(\theta)\frac{f(\mathbf{x};\theta)}{\int_{-\infty}^{\infty} f(\mathbf{x};\theta^\#))d\theta^\#}}{\int_{\theta^*} \pi_*(\theta^*)\frac{f(\mathbf{x};\theta^*)}{\int_{-\infty}^{\infty} f(\mathbf{x};\theta^\#))d\theta^\#}d\theta^*} \qquad (20)$$

what can be simplified to

$$\pi_*(\theta|\mathbf{x}) = \frac{\pi_*(\theta)f(\mathbf{x};\theta)}{\int_{\theta^*} \pi_*(\theta^*)f(\mathbf{x};\theta^*)d\theta^*} \qquad (21)$$

The latter equation is the same as the original equation (3) for the Bayesian inference. Only one prior distribution, now $\pi_*(\theta)$, influences the estimation; the 1$^{st}$ flat prior is removed. The only difference between (3) and (21) is the notation for the unconditional prior and conditional posterior distribution with $\pi$ and $\pi_*$. This means, that the updating by (19) is reasonable.

This result for (19) also means that the Bayesian inference implies a combination of a prior confidence distribution for the sample $x$ with a prior information which does not depend on $x$. There is no restriction that prior confidence distribution $\pi(\theta|x)$ in (19) must be the result of a Bayesian inference. Therefore, the updating of a confidence distribution $\pi(\theta|x)$ by additional information according to equation (19) for generating a posterior confidence distribution $\pi_*(\theta|x)$ can be applied in Bayesian and non-Bayesian inference. This is an extension of the Bayesian approach (4) to non-Bayesian inference and is also consistent with Lindsey's (1996) understanding of Bayesian inference as a kind of penalized likelihood estimation. The point estimation for the posterior confidence distribution $\pi_*(\theta|x)$ of equation (19) is computed according to equation (4). Its median is also a reasonable point estimator.

The mechanism of the new approach can be demonstrated by the example of the normal distribution, now with actual prior information according to Figure 6c. The resulting Bayesian and non-Bayesian posterior confidence distributions again are the same (Figure 6d).

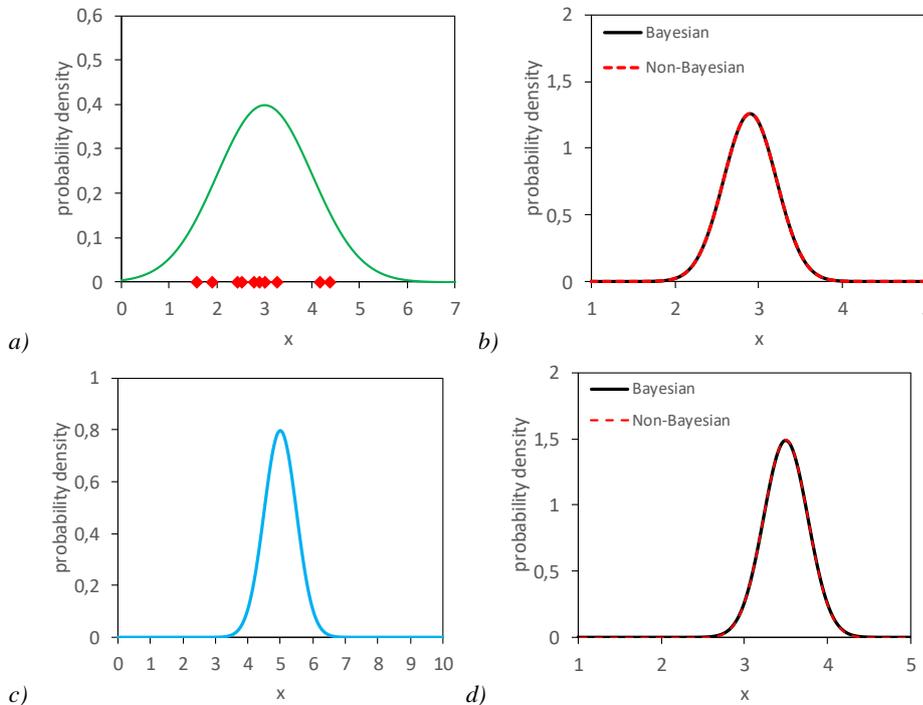

*Figure 6: Example of confidence distributions according to Bayesian and non-Bayesian inference: a) samples and actual normal distribution, b) confidence distribution of non-Bayesian and Bayesian inference with flat prior distribution, c) non-flat prior distribution and d) resulting confidence distribution for the Bayesian and the new non-Bayesian approach.*

The asymptotic behaviour of the new point estimation is also shown for the example of upper bound magnitude in section 2.2, Figure 2b. The estimation is $\hat{\Delta}= 0.5$ for (8) and an exponential prior confidence distribution is applied according to section 3.1. The dependence between point estimation and standard deviation of the prior information is shown in Figure 7. The behaviour equals the behaviour for the Bayesian estimation for the expectation of a normal distribution in Figure 2a. If the standard deviation of the prior is very small, the point estimation equals the expectation of the prior distribution. If the standard deviation of the prior is very high (no informative case) the point estimation equals the expectation of the prior confidence distribution from non-Bayesian inference.

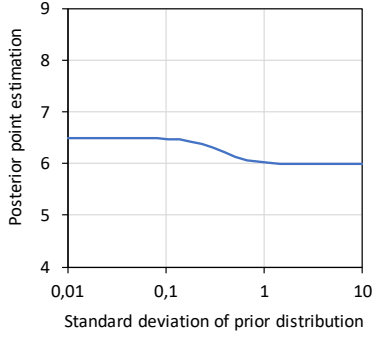

*Figure 7: Example of the relation between new point estimation of upper bound magnitude with a prior confidence distribution and the standard deviation of the normal prior distribution.*

## 3.3 Further estimation methods with prior information

Now further alternatives are developed. For fixed $m_{max}$ and sample size $n$, the CDF $F_{obs}$ of random variable $m_{max}^{obs}$ is according to the extreme value statistics (cf. Beirlant et al. 2004, keywords block maximum and sample maximum)

$$F_{obs}(m_{max}^{obs}|m_{max}) = [F(m_{max}^{obs}|m_{max})]^n. \qquad (22)$$

The CDF is here a conditional one with random parameter $m_{max}$ as condition. The corresponding conditional PDF is the first derivative of (22)

$$f_{obs}(m_{max}^{obs}|m_{max}) = f(m_{max}^{obs}|m_{max})[F(x|m_{max})]^{n-1} \qquad (23)$$

The entire population of $m_{max}^{obs}$ with the entire population of random parameter $m_{max}$ (with PDF $\pi(m_{max})$) as background has the independent PDF

$$f_{obs}(m_{max}^{obs}) = \int_{m_{max}} f(m_{max}^{obs}|m_{max})[F(m_{max}^{obs}|m_{max})]^{n-1} \pi(m_{max}) dm_{max} \qquad (24)$$

Integration of (24) results in the CDF $F_{obs}$ for $m_{max}^{obs}$ and a new simple estimator for $m_{max}$ is

$$\hat{m}_{max} = \Pi^{-1}\left(F_{obs}(m_{max}^{obs})\right). \qquad (25)$$

The inverse function is of CDF $\Pi$ of the random parameter $m_{max}$. The formulation is general. The solution for (25) depend on the distribution of $m_{max}$ - the prior information. The solution can be also and is here computed numerically. In the current paper, the CDF $F$ is the truncated exponential distribution with their parameter $m_{max}$ and $\beta$ (equation (1)). The estimation by (25) is a transformation function $m_{max}^{obs} \rightarrow \hat{m}_{max}$ via the CDF $F_{obs}$ and the inverse CDF $\Pi^{-1}$ of the unconditional prior distribution. Examples illustrate the principal. The upper bound magnitude has the CDF of a symmetric beta distribution (s. appendix) in the range between $m$=3 and 14 with expectation 7. Different standard deviations are considered. The scale parameter of (1) is $\beta = \ln(10)$ and sample sizes $n$=1000, 4000 and 16000 are considered. The CDFs of $m_{max}^{obs}$ are shown in Figure 8a. The corresponding transformation functions for the estimation are shown in Figure 8b. The lower range of this function is a straight line for numerical reasons. There is no simple opportunity to quantify the corresponding estimation error and/or the confidence distribution for this estimation.

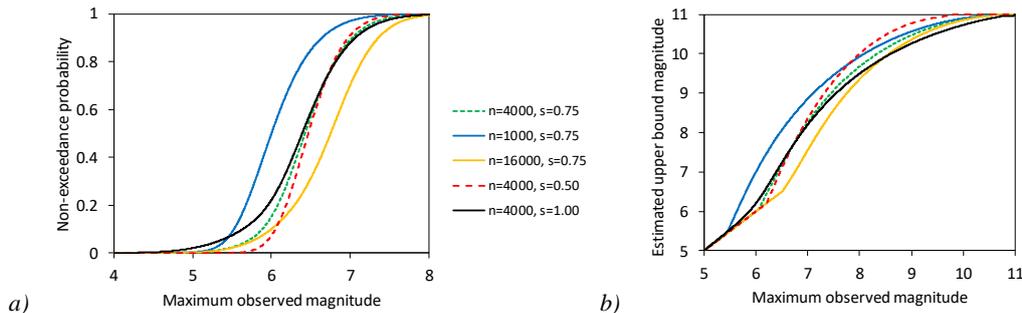

*Figure 8: Examples for the approach according to (25): a) CDF $F_{obs}$ and b) corresponding transformation functions according to (25) for different sample sizes $n$ and standard deviations $S$ of the underlaying distribution of $m_{max}$.*

Consideration of prior information in the inference for the upper bound earthquake magnitude - submitted major revisionThere are further opportunities to estimate the upper bound with this approach. The perspective is changed for this purpose and $m_{max}^{obs}$ is fixed in the conditional PDF $f_{obs}(m_{max}^{obs}|m_{max})$ of (23). The conditional distribution $\pi(m_{\max}|m_{max}^{obs})$ can be formulated according to the well-known rules for bivariate distributions (Bayes theorem for continuous random variables) with the unconditional PDF $f_{obs}(m_{max}^{obs})$ of (24) and PDF $\pi(m_{max})$:

$$\pi(m_{\max}|m_{max}^{obs}) = \frac{f_{obs}(m_{max}^{obs}|m_{max})\pi(m_{max})}{f_{obs}(m_{max}^{obs})}. \tag{26}$$

The conditional expectation $E[m_{\max}|m_{max}^{obs}]$ of PDF (26) is a point estimator for $m_{max}$ under the assumption that a finite expectation exists. It is called here new expectation estimator (method). The median of (26) can also be used as point estimation and is called new median estimator (method). The standard deviation can also be computed for this posterior distribution and quantifies the standard error. The new approach (26) is similar to the classical Bayesian estimation; $\pi(m_{\max})$ is a kind of prior distribution and $\pi(m_{\max}|m_{max}^{obs})$ is a kind of conditional posterior distribution. Therefore, the same notation is used as for the Bayesian inference. But there are also differences. For the new approach (26), there must be an actual and known underlying distribution for the upper bound, and there is no explicit relation to the maximum likelihood method and function. Moreover, the distributions $f_{obs}(m_{max}^{obs}|m_{max})$ and $f_{obs}(m_{max}^{obs})$ are explicate parts of (26) and not explicit parts of the Bayesian inference for $m_{\max}$ according to (3).

## 4 Performance of the inference approaches

### 4.1 Concept of the performance evaluation

The performance of the different approaches is validated by a numerical study. Therein the estimated parameter $m_{max}$ is a random parameter. Its distribution is defined and represents the population of $m_{max}$. For every repetition of a Monte Carlo simulation, the current value of the $m_{max}$ is simulated by a random generator. Then the observed maximum magnitude $m_{max}^{obs}$ is also simulated by a random generator under consideration of the current $m_{max}$ and other parameters, for a fixed sample size. The defined underlying distribution of $m_{max}$ is also the prior distribution. The prior information is correct in this way. The $\beta$-value is fixed and known in the estimations. These are computed with $m_{max}^{obs}$ according to the different inference methods (with or without prior information). Finally, two large samples of estimations are generated of size $n = 1000$ and can be analysed. One sample includes the point estimations $\hat{m}_{max}$, the second includes of the estimated non-exceedance probability $\hat{F}_{con}(m_{max})$ of the final confidence distribution for the actual $m_{max}$.

The performance is quantified by various measures. A good performance implies a small mean squared error (MSE) being the expectation $E[(\hat{m}_{max} - m_{max})^2]$ and is estimated by the mean of the sample of $(\hat{m}_{max} - m_{max})^2$ from the Monte Carlo simulation. The MSE is the most important and popular measure for the evaluation of a point estimation (cf. Lindsey 1996, Beirlant et al. 2004). The absolute value of the differences $E[\hat{m}_{max}] - m_{max}$ is the bias, which should also be small and is computed via the mean of the sample of $\hat{m}_{max}$ from the simulation.

A certain correlation between the error $\hat{m}_{max} - m_{max}$ and $m_{max}$ is expected for the Bayesian estimation with actual prior information. This implies a kind of local bias and it means a bias that depends on actual $m_{max}$. Estimations without actual prior information should be less effected or not effected at all by such correlation and local bias.

The results of the estimated confidence distribution are presented by the sample of $\hat{F}_{con}(m_{max})$. This should have a sample mean near 0.5 and a variance near 1/12 since the estimations should follow a uniform distribution (Johnson et al. 1995) between 0 and 1 as every univariate CDF without probability mass does. Additionally, the probability, that the actual value of $m_{max}$ falls inside an estimated confidence interval with significance level α, should be equal to 1-α (cf. Lindsey 1996). This is not ensured if $\hat{F}_{con}(m_{max})$ of the actual value of $m_{max}$ is correlated with this. This instance is expected for estimations with prior information.

Plots are also used to validate the performance visually. The plot of the point estimations shows the relation between the actual value of $m_{max}$ and the error being the difference $\hat{m}_{max} - m_{max}$. This complies with the residual plot of the regression analysis (Fahrmeir et al. 2013).

For the confidence distribution, the empirical distribution function is applied, which has an estimated CDF

$$\hat{F}(x_i) = \frac{i}{n+1} \tag{27}$$

for each observation $x_i$ of the ordered sample $x_1 \leq x_2 \leq \cdots \leq x_i \leq \cdots \leq x_n$ of size $n$. Here the ordered sample of $\hat{F}_{con}(m_{max})$ is compared with the corresponding empirical CDF of (27). These values can also be transformed by the inverse CDF of the standard normal distribution and a Q-Q plot with standard normal margins is generated. The tails of the confidence distribution can be better validated with such a plot and many researchers are more familiar with the Q-Q plot for normal distributions.

Before the concept is applied to the estimation of the upper bound magnitude, it is applied to the very simple estimation of the expectation of a normally distributed random variable $X$ in the following section.

### 4.2 Performance of the expectation estimation for a normal distribution

The expectation $E[X] = \mu$ of a normally distributed random variable $X$ with PDF (6) is estimated in this performance study with 10000 repetitions. The variance of $X$ is known with $V[X] = \sigma^2 = 1$, the sample size is $n = 10$. The non-Bayesian point estimator is the sample mean $\bar{x}$ as well as the moment estimator and the ML estimator. The corresponding confidence interval is modelled by a normal distribution with expectation $\mu_{con} = \bar{x}$ and variance $\sigma_{con}^2 = 1/n$ (Johnson et al. 1994). The distribution of the expectation of $X$, which is a normal distribution, is also the prior information. Its parameters are $\mu_{pri} = 0$ and $\sigma_{pri}^2 = 0.4^2$.

As might be expected, the applied inference methods perform perfectly. The average of the error $\hat{E}[X] - E[X]$ is very near to 0. The MSE for the entire population is only 0.06 for the version with prior information. The estimation without prior information has a higher MSE with 0.10. However, the better overall performance of the method with prior information is paid by a local bias according to Figure 9a. The estimation error correlates strongly with the actual expectation; the estimated Pearson's correlation coefficient is $\hat{R} = 0.627$. The correlation for the inference without prior information is not significant, with $\hat{R} = -0.012$.

Furthermore, the modelled confidence intervals also perform well from the global point of view. The plots of Figure 9b and c are almost perfectly in line with the ideal. This corresponds to the mean of the sample of $\hat{F}_{con}(E[X])$, which is very close to ½ and the variance is very close to 1/12. The correlation between actual expectation $E[X]$ and $\hat{F}_{con}(E[X])$ is $\hat{R} = 0.012$ for the inference without prior information and is $\hat{R} = 0.606$ for the case with prior information. The latter implies a bias. A simple analysis illustrates the effect of this weakness. The share of estimations $\hat{F}_{con}(E[X])$ over the 50% quantile with $\hat{F}_{con}(E[X]) > 0.5$ should be around 50%. The empirical share is computed for the generated data and listed in **Fehler! Verweisquelle konnte nicht gefunden werden.**. The analysis is separately performed for the cases $E[X] \leq 0$ and $E[X] > 0$. The empirical shares of estimations with prior information differ considerably from the correct value of 50%. The absolute differences are around 20%. In contrast, the empirical share of estimations without prior information is very close to the correct values in all cases. These local biases are the price for the smaller global MSE of the point estimations with prior information. In summary, the concept for the evaluation of the performance is positively validated by the example of expectation of a normal distribution.

*Table 1: Share of estimations $\hat{F}_{con}(E[X]) > 0.5$*

| Case | Correct share | Empirical share with prior | Empirical share without prior |
|---|---|---|---|
| $E[X] \leq 0$ | 50.0% | 30.4% | 50.1% |
| $E[X] > 0$ | 50.0% | 72.5% | 50.7% |

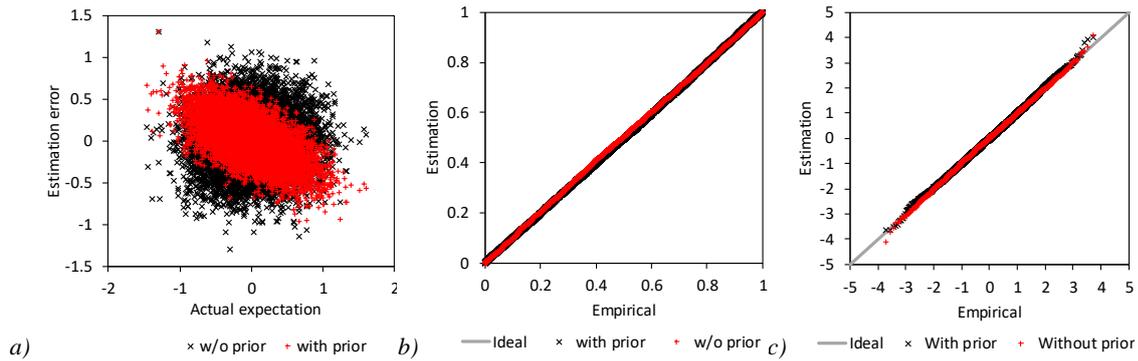

*Figure 9: Performance of the estimation methods for the expectation of a normal distribution: a) estimation errors versus actual expectation, b) plot of estimated and empirical non-exceedance probabilities from the modelled confidence regions, c) transformation of c) to a Q-Q plot for normal margins.*

### 4.3   Performance of the inference for the upper bound magnitude

The following situation is constructed for the performance study for the estimation of the upper bound magnitude with 1000 repetitions. Every parameter variant is computed by the same initialisation of the random number generator. This means that the underlying random numbers are the same. The known $b$-value is 1 with corresponding $\beta = ln(10)$. The observation threshold is $m_{min} = 2.5$. It is drawn out that the following results also hold for other values of $m_{min}$ if the corresponding values of $m_{max}$ and $\hat{m}_{max}$ are equivalently shifted. The standard analysis is done for sample size $n = 4000$ and a prior distribution with expectation $E[m_{max}] = 6.5$ and standard deviation $S[m_{max}] = 0.75$. A symmetric beta distribution is applied for the population of $m_{max}$ and the prior distribution. Its lower bound is $m_{min}$, its upper bound is 10.5. The beta distribution is applied since it has a PDF similar to the normal distribution (Figure 3a) and a normal distribution is frequently used as prior for upper bound magnitudes. However, a simulation with a normal distributed $m_{max}$ can lead to the impossible case $m_{max} < m_{min}$. Therefore, an approximation for the normal distribution is applied. Besides the standard variant with $S[m_{max}] = 0.75$ and $n = 4000$, further analyses are done for sample sizes $n = 1000$ and $n = 16000$ as well as for standard deviations $S[m_{max}] = 0.50$ and $S[m_{max}] = 1.00$.

The following point estimators are applied: The classical Bayesian method (3-4), the bias free method (12) of Pisarenko (1991), the constructed estimator (14) of Raschke (2012), and their variants with prior information and an exponential prior confidence distribution (section 3.2). The new methods of section 3.3 are also applied: new simple with (25), new expectation and new median method with (26). Following confidence distributions are analysed: Bayes according to (3), exponential for the bias free estimator with and without prior information, constructed estimator (14) with and without prior, and the new approach (26) which considers prior information. Pisarenko's (1991) proxy (13) also is considered. The estimation method (10) is not considered because of the problem of iteration (Figure 5).

Plots of the estimation errors are only presented for the standard variant in Figure 10. The ideal is a line according to Figure 10a. The bias free estimator of Pisarenko (Figure 10b) has clearly the smallest local bias. The constructed estimator (Figure 10c) has considerable local bias. But the highest is found in the classical Bayesian inference (Figure 10d). The combination of previous non-Baysian inference with prior information does not lead to a significant improvement (Figure 10e and f). The results for new estimators of section 3.3 are presented by Figure 10g to i. The simple variant is slightly similar to the bias free estimator. The equivalence between the new expectation method and the classical Bayesian estimation is surprising, and the new median method has a smaller local bias than the new expectation method.

The global biases are listed in Table 2. The classical Bayesian approach, the new methods and the unbiased estimator have lowest global biases. The global bias of the constructed method and the combination of prior information with the non-Bayesian methods are higher. A larger sample size reduces this bias and a smaller standard deviation of the prior distribution too. A similar influence of these parameters is stated for the MSE in Table 3. The smallest global MSE has the new median estimator, followed by the classical Bayesian and the new expectation method with equal values. The constructed method has a higher MSE, which is not improved by a prior distribution. The unbiased method has the highest global MSE. However, they decrease by increasing the sample size most rapidly, and the improvement by prior information is significant. The correlation coefficients in Table 4 broadly confirm the statements about the plots and the corresponding local bias.

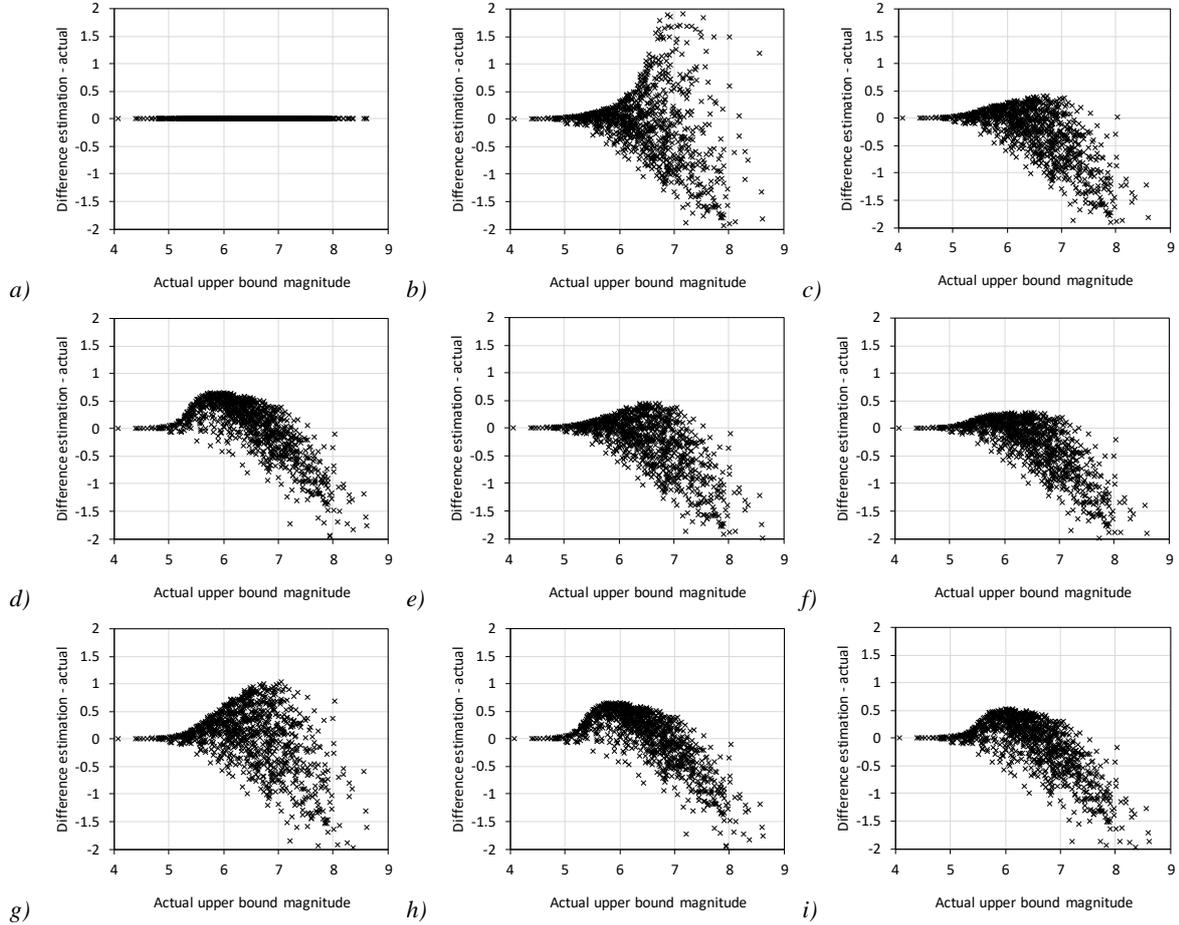

*Figure 10: Performance of the estimation methods for upper bound magnitude for the standard situation: a) ideal relation, b) bias-free estimator, c) constructed estimator, d) Bayes estimator, e) bias-free estimator with prior, f) constructed estimator with prior, g) new-simple, h) new-expectation and i) new-median ($S[m_{max}] = 0.75$ and $n = 4000$).*

*Table 2: Bias of the different estimation methods for the upper bound magnitude.*

| Standard deviation of prior | Sample size n | Estimation method | | | | | | | |
|---|---|---|---|---|---|---|---|---|---|
| | | Unbiased | Constr. | Bayes | Unbiased & Prior | Constr. & Prior | New simple | New expec. | New median |
| 0.75 | 4000 | -0.0599 | -0.2886 | 0.0070 | -0.2757 | -0.3126 | -0.0019 | 0.0076 | -0.0887 |
| 0.75 | 1000 | -0.1170 | -0.6281 | 0.0163 | -0.4912 | -0.5804 | 0.0077 | 0.0167 | -0.0446 |
| 0.75 | 16000 | -0.0268 | -0.0931 | -0.0007 | -0.1269 | -0.1357 | -0.0053 | -0.0000 | -0.0647 |
| 0.5 | 4000 | -0.0314 | -0.2153 | 0.0076 | -0.2177 | -0.2342 | 0.0026 | 0.0081 | -0.0502 |
| 1 | 4000 | -0.0984 | -0.3723 | 0.0057 | -0.3354 | -0.3922 | -0.0042 | 0.0063 | -0.1085 |

*Table 3: MSE of the different estimation methods for the upper bound magnitude*

| Standard deviation of prior | Sample size n | Estimation method | | | | | | | |
|---|---|---|---|---|---|---|---|---|---|
| | | Unbiased | Constr. | Bayes | Unbiased & Prior | Constr. & Prior | New simple | New expec. | New median |
| 0.75 | 4000 | 0.8494 | 0.3577 | 0.2580 | 0.3539 | 0.3673 | 0.3174 | 0.2580 | 0.1853 |
| 0.75 | 1000 | 2.4702 | 0.8956 | 0.4498 | 0.7719 | 0.8108 | 0.6265 | 0.4499 | 0.3028 |
| 0.75 | 16000 | 0.2241 | 0.1180 | 0.1053 | 0.1278 | 0.1312 | 0.1176 | 0.1053 | 0.0784 |
| 0.5 | 4000 | 0.4892 | 0.2292 | 0.1607 | 0.2224 | 0.2192 | 0.1974 | 0.1607 | 0.1111 |

| | | | | | | | | |
|---|---|---|---|---|---|---|---|---|
| 1 | 4000 | 1.3137 | 0.5434 | 0.3691 | 0.5243 | 0.5591 | 0.4691 | 0.3691 | 0.2681 |

*Table 4: Correlation between actual upper bound magnitude and estimation error.*

| Standard deviation of prior | Sample size n | Estimation method | | | | | | | |
|---|---|---|---|---|---|---|---|---|---|
| | | Unbiased | Constr. | Bayes | Unbiased & Prior | Constr. & Prior | New simple | New expec. | New median |
| 0.75 | 4000 | -0.1269 | -0.6757 | -0.6893 | -0.6464 | -0.7399 | -0.4084 | -0.6896 | -0.6692 |
| 0.75 | 1000 | -0.1410 | -0.7593 | -0.8940 | -0.6999 | -0.8144 | -0.5560 | -0.8941 | -0.8795 |
| 0.75 | 16000 | -0.1154 | -0.5130 | -0.4525 | -0.5640 | -0.6262 | -0.2770 | -0.4527 | -0.5380 |
| 0.5 | 4000 | -0.1006 | -0.5774 | -0.8144 | -0.6264 | -0.7292 | -0.4845 | -0.8146 | -0.7828 |
| 1 | 4000 | -0.1574 | -0.7310 | -0.6127 | -0.6685 | -0.7623 | -0.3701 | -0.6129 | -0.6329 |

The Q-Q plots with standard normal margins (cf. sections 4.1 and 4.2) for the confidence distributions are shown in Figure 11. It is clear that the approximation of the confidence distribution for the unbiased and construction estimator by an exponential distribution fails. The improvement by considering the prior distribution is limited. Conversely, the confidence distributions of the Bayesian method, the new method and the approximation by (13) perform well from the global point of view. The values for the sample mean and the sample variance of $\hat{F}_{con}(m_{max})$ in Table 5 and Table 6 confirm this. The results for the classical Bayesian inference and the new expectation method with (26) are very similar; the small differences could be due to numerical imperfections. The proxy (13) has the same value for every variant of parameters. The reason for this behaviour is the same initialisation of the random generator in the Monte Carlo simulations. In addition, the proxy (13) has an insignificant correlation (Table 7). The absolute correlation values of the Bayesian and all new methods are once again very high and indicate a poor behaviour of their (posterior) confidence distribution.

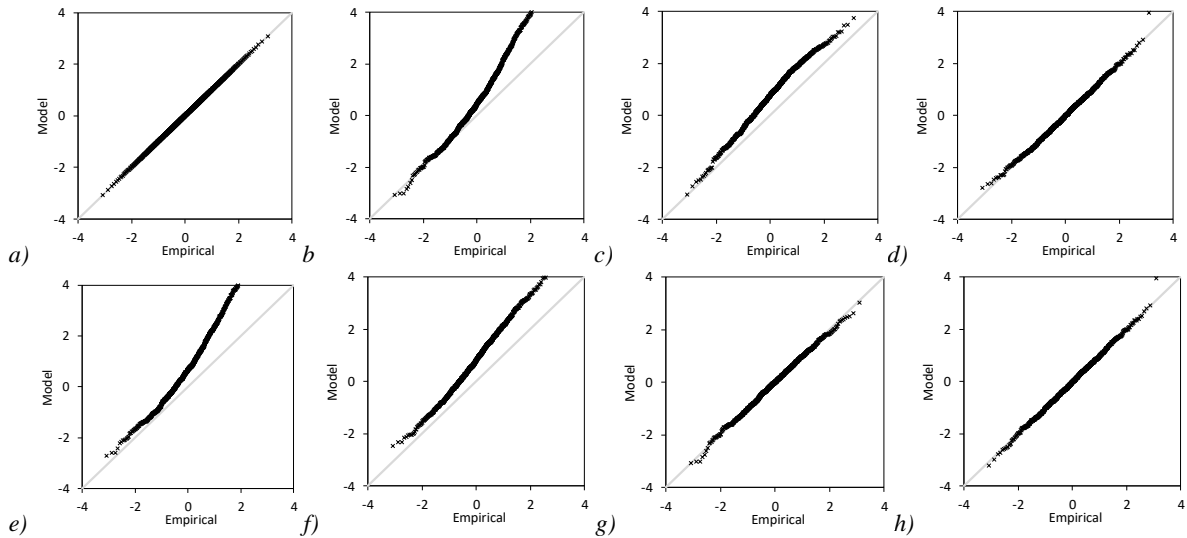

*Figure 11: Q-Q plot with normal margins for the non-exceedance probabilities of the actual upper bound magnitude according to the modelled confidence intervals for the standard situation: a) ideal, b) for the bias-free estimator, c) constructed estimator, d) Bayes estimator, e) bias-free estimator with prior, f) constructed estimator with prior (equivalent to approximation and prior) g) approximation and h) the new method.*

*Table 5: Mean value of exceedance probability of the actual upper bound magnitude according to the estimated (posterior) confidence distribution.*

| Standard deviation of prior | Sample size n | Estimation method | | | | | | |
|---|---|---|---|---|---|---|---|---|
| | | Unbiased | Constr. | Bayes | Unbiased & Prior | Constr. & Prior | Proxy | New |
| 0.75 | 4000 | 0.6137 | 0.6817 | 0.4931 | 0.6589 | 0.6905 | 0.4973 | 0.4908 |
| 0.75 | 1000 | 0.5636 | 0.5865 | 0.5037 | 0.6109 | 0.6214 | 0.4973 | 0.4902 |
| 0.75 | 16000 | 0.6137 | 0.6817 | 0.4931 | 0.6589 | 0.6905 | 0.4973 | 0.4946 |
| 0.5 | 4000 | 0.6095 | 0.6768 | 0.4929 | 0.6536 | 0.6790 | 0.4973 | 0.4919 |

| | 1 | 4000 | 0.6184 | 0.6875 | 0.4958 | 0.6640 | 0.6996 | 0.4973 | 0.4903 |

*Table 6: Sample variance of exceedance probability of the actual upper bound magnitude according to the estimated (posterior) confidence distribution.*

| Standard deviation of prior | Sample size n | Estimation method | | | | | | |
|---|---|---|---|---|---|---|---|---|
| | | Unbiased | Constr. | Bayes | Unbiased & Prior | Constr. & Prior | Proxy | New |
| 0.75 | 4000 | 0.1085 | 0.0860 | 0.0839 | 0.0998 | 0.0887 | 0.0825 | 0.0846 |
| 0.75 | 1000 | 0.1092 | 0.0658 | 0.0839 | 0.0946 | 0.0735 | 0.0825 | 0.0841 |
| 0.75 | 16000 | 0.1017 | 0.0900 | 0.0816 | 0.0963 | 0.0905 | 0.0825 | 0.0840 |
| 0.5 | 4000 | 0.1084 | 0.0810 | 0.0845 | 0.0989 | 0.0863 | 0.0825 | 0.0848 |
| 1 | 4000 | 0.1087 | 0.0899 | 0.0830 | 0.0990 | 0.0888 | 0.0825 | 0.0845 |

*Table 7: Correlation between actual upper bound magnitude and exceedance probability of the actual upper bound magnitude according to the estimated (posterior) confidence distribution.*

| Standard deviation of prior | Sample size n | Estimation method | | | | | | |
|---|---|---|---|---|---|---|---|---|
| | | Unbiased | Constr. | Bayes | Unbiased & Prior | Constr. & Prior | Proxy | New |
| 0.75 | 4000 | 0.2801 | 0.5243 | 0.6266 | 0.4868 | 0.6041 | 0.0550 | 0.6365 |
| 0.75 | 1000 | 0.2972 | 0.5734 | 0.8634 | 0.5609 | 0.6913 | 0.0550 | 0.8648 |
| 0.75 | 16000 | 0.2321 | 0.3913 | 0.3363 | 0.3641 | 0.4466 | 0.0550 | 0.3835 |
| 0.5 | 4000 | 0.2207 | 0.4471 | 0.7688 | 0.5266 | 0.6491 | 0.0551 | 0.7699 |
| 1 | 4000 | 0.3289 | 0.5695 | 0.5235 | 0.4675 | 0.5838 | 0.0549 | 0.5526 |

# 5 Conclusion and Discussion

In this paper, the opportunities to consider prior information for the estimation of the upper bound magnitude was researched and extended. One contribution is the extension of the Bayesian inference approach to non-Bayesian inference by replacing of the likelihood function in (3) by a prior confidence distribution according to section 3.2. This extension is universal as shown for the estimation of the expectation of a normal distribution. It also leads to an acceptable convergence of the point estimation of the upper bound magnitude (Figure 7) in contrast to the behaviour of classical Bayesian point estimation of the upper bound (Figure 2).

However, the reasonable approximation of the confidence distribution for non-Bayesian methods by an exponential distribution does not perform well according to the study in section 4.3.

The here derived new (posterior) confidence distribution (26) of section 3.3 results in the (almost) same estimation as the classical Bayesian inference. The advantage of the new approach is that some restrictions for the classical Bayesian inference do not apply. The estimations do not need to converge to the maximum likelihood estimation if the information content of the prior distribution decreases. Also, the criticism by Holschneider et al. (2011) on the Bayesian inference for the upper bound magnitude does not apply to the new methods. It is assumed that the new confidence distribution (26) equals exactly the Bayesian posterior distribution (3) even though it could not be analytically proven here. The new (26) and the Bayesian approach could have a similar relation as the moment method and the likelihood method for the expectation of a normal distribution: different approaches with equivalent results. A further interesting result is that the median of (26) is a better point estimator (smaller MSE) than the expectation of (26).

The new approach of section 3.3 provides a further opportunity to estimate the upper bound magnitude by (25). Its local bias is also smaller than the one of all other estimations with prior information with smallest absolute value of correlation in Table 4.

The bias free estimator of Pisarenko (1991, here equation (12)) has the smallest absolute value of correlation for the point estimation of all considered methods in section 4.3. The consideration of prior information also reduces the global MSE of this estimator significantly for the price of a higher local bias. The constructed estimator (14) of Raschke (2012) is a modification of the confidence distribution by Pisarenko (1991, here (13)) and is not improved by prior information.

Correspondingly, the idea of Kijko and Smit (2014) does not work. However, the confidence distribution by Pisarenko (1991, here (13)) performs very well (Figure 11, Table 5, Table 6) in contrast to the other confidence distributions without prior information. It also practically has no local bias since the absolute correlation is small (Table 7); the only shortcoming is the probability mass for $m_{max} = \infty$. Correspondingly, the possibility to derive a point estimator are limited. Other acceptable confidence distributions are from the classical Bayesian inference and from the new approach (26) with equivalent results. Their weakness is the local bias with high absolute value of correlation (Table 7).

In summary, none of the researched and/or developed estimation methods can be recommended without compromise. A reduction of the MSE by a prior distribution is paid with a local bias; smaller upper bound magnitudes are overestimated, and higher upper bounds are underestimated. Only very small magnitudes and medium size magnitudes have no bias. A study about the consequences for a PSHA could be a support for modeller's decision in a PSHA concerning the estimation of the upper bounds. This would be in the sense of the risk perspective of the extended version of classical Bayes estimation (Lindsey, 1996).

Further research is recommended. For example, further estimation methods, parameter constellation and types of prior distributions could be considered. The basic design for such performance analysis was provided in section 4.1 and validated in section 4.2. Such a performance study is recommended for the parameterisation of every important PSHA (e.g. for building codes or nuclear facilities).

It may also be that the confidence distribution by Pisarenko (1991, here (13)) can be modified in a more sophisticated way than done by Raschke (2012) to eliminate the probability mass for $m_{max} = \infty$. In addition, the issue of the basic approach by Kijko and Graham (1998) and Kijko (2004), which was explained in section 3.1, should be clarified. The improvement of inference methods that does not need prior information is generally recommended because of the bias of inference with prior information (Figure 9a, Figure 10, Table 4 and Table 7). Moreover, prior information includes additional uncertainties according to sections 2.3 and 2.4. There is a concrete example for this issue, the results by Coppersmith (1994) respectively Johnston et al. (1994). They suggested parametrized normal distributions as prior distributions for different tectonic regimes. The assumption about the distribution type was not validated even though it influences the estimations (sections 2.3). Furthermore, the suggested parametrization and the corresponding parameter estimation by Coppersmith (1994) could not be reproduced by the author of the current paper. And as-far-as the author knows, there is not a method for the extraction of prior information about the upper bound magnitude which was published and discussed in and accepted by the scientific community of mathematical statisticians and/or validated by extensive numerical researches. Additionally, any uncertainty and error of a prior distributions propagates to all estimations for the upper bound magnitude which consider this prior (section 2.4).

An underestimation of the upper bound magnitude leads to an underestimation of seismic hazard. In this case it is worse at the first glance. However, every overestimation of a type of hazard leads to a relative underestimation of other types of hazard and the resources for retrofitting could be incorrectly managed.

There are further issues which affect the estimation of the upper bound magnitude. The measurement error of the magnitudes was already mentioned in the introduction. Moreover, the basic assumption of a truncated exponential distribution (Gutenberg-Richter relation) is also not verified in many researches. There are arguments indicating that this is an insufficient simplification (Raschke 2014b). This should also be considered in future research and PSHA.

# Acknowledgement
The author thanks the referees for several constructive remarks and suggestions which were considered in the revisions of this paper.

Consideration of prior information in the inference for the upper bound earthquake magnitude - submitted major revision

# Appendix

The PDF of a shifted log-normal distribution (Johnson et al. 1994) is with $\sigma > 0$ and $\mu > 0$

$$f(x) = \begin{cases} \frac{1}{\sqrt{2\pi}\sigma x} exp\left(-\frac{(\ln(x-\alpha)-\mu)}{2\sigma^2}\right) & if\ x > \alpha \\ 0 & if\ x \leq \alpha \end{cases}.$$

Its expectation is
$$E[X] = \alpha + exp(\mu + \sigma^2/2),$$
and its variance is
$$V[X] = exp(2\mu + \sigma^2)(exp(\sigma^2) - 1).$$

The PDF of a shifted and scaled beta distribution (Johnson et al. 1995) is with $a < b$ and $p > 0$ and $q > 0$

$$f(x) = \begin{cases} \frac{x^{p-1}(1-x)^{q-1}}{B(p,q)(b-a)} & if\ x \geq a\ and\ x \leq b \\ 0 & if\ x < \alpha\ and\ x > b \end{cases}.$$

Its expectation is
$$E[X] = a + (b-a)\frac{p}{p+q},$$
and its variance is
$$V[X] = (b-a)^2 \frac{pq}{(p+q+1)(p+q)^2}.$$

The beta distribution is symmetric if $p = q$.